\documentclass[a4paper,11pt]{article}
\pdfoutput=1 % if your are submitting a pdflatex (i.e. if you have
\usepackage{jcappub} % for details on the use of the package, please
\usepackage[T1]{fontenc} % if needed

\usepackage{bm} % to write vectors as bold letters in math
\newcommand{\im}{\text{Im\,}}
\newcommand{\re}{\text{Re\,}}
\usepackage{braket}% bra ket typesetting

\title{\boldmath A strategy for optimal material identification in solar dark photon absorption}

\author[a,b]{Theresa M.~Backes,}
\author[a]{Riccardo Catena}
\author[b]{and Michael Kr\"amer}

\affiliation[a]{Department of Physics, Chalmers University of Technology, SE-412 96 G\"oteborg, Sweden}
\affiliation[b]{Institut f\"ur Theoretische Teilchenphysik und Kosmologie, RWTH Aachen University, D-52056 Aachen, Germany}

\emailAdd{theresa.backes@chalmers.se}
\emailAdd{catena@chalmers.se}
\emailAdd{mkraemer@physik.rwth-aachen.de}

\abstract{Dark photons with masses in the 1–100 eV range can be produced in the Sun and subsequently absorbed in terrestrial detectors, offering a promising avenue for probing hidden-sector physics beyond the Standard Model. In this work, we develop a theoretically grounded strategy to identify optimal detector materials for solar dark photon absorption. Our strategy builds on a material-independent upper limit on the absorption rate, which we derive from Kramers–Kronig relations applied separately to the longitudinal and transverse dark photon modes. We show how the optimal material properties depend on the dark photon mass relative to the detector's plasma frequency, identifying the conditions under which a detector can saturate the theoretical upper limit. We then assess the performance of commonly used detector materials in light of these criteria and comment on the prospects of metamaterials featuring tunable plasma frequencies. Our results provide a general and model-independent framework to effectively guide the design of next-generation experiments targeting solar dark photons.}

\begin{document}
\maketitle
\flushbottom

\section{Introduction}\label{sec:intro}
A possible explanation for the current lack of discovery of Weakly Interacting Massive Particles (WIMPs) in direct detection experiments is that dark matter (DM) may be lighter than the nucleons bound in the nuclei that make up detector materials, and therefore too light to induce an observable nuclear recoil in existing setups~\cite{Essig:2011nj}. Such light, i.e. sub-GeV, DM particles can be in thermal equilibrium in the early Universe, achieve the correct cosmological abundance, and remain compatible with the Lee–Weinberg bound if the Standard Model (SM) of particle physics is extended by an additional force carrier that mediates the interaction between DM and known species, and that is responsible for both thermal equilibrium and the present cosmological density of the DM component of our Universe~\cite{PhysRevLett.39.165}. 

The so-called ``dark photon'' -- a gauge boson associated with an additional $U(1)$ symmetry beyond the SM -- is the archetypal example of such a new force carrier~\cite{HOLDOM1986196,FAYET1990743,FAYET1980285,FAYET1981184}. Contrary to the SM photon, the dark photon can acquire a mass, either through the Higgs mechanism or by introducing a St\"uckelberg field into the theory~\cite{Stueckelberg, Ruegg:2003ps}. 

The search for massive dark photons relies on different methodologies, depending on whether the dark photon is heavier or lighter than twice the electron mass, $m_e$~\cite{Fabbrichesi:2020wbt,PhysRevD.104.095029}. Dark photons lighter than $2 m_e$ do not decay visibly and are primarily searched for in atomic and nuclear experiments that aim to detect modifications of the Coulomb force~\cite{PhysRevD.2.483}; in light-shining-through-a-wall experiments~\cite{EHRET2010149}; in deviations from the black-body spectrum of the cosmic microwave background~\cite{Mirizzi:2009iz}; through anomalous energy-transport effects in stars~\cite{Redondo_2013,Pospelov2}; and, finally, via their absorption in terrestrial detectors~\cite{Redondo_2008,Pospelov1}. 

Solar dark photon absorption in terrestrial detectors is the experimental methodology at the core of the present work. In particular, we are interested in developing a theoretically driven approach to identify optimal detector materials for solar dark photon absorption. We believe that addressing this question could guide the design of next-generation dark photon search experiments in a systematic and effective manner.

Specifically, we propose a strategy to identify optimal detector materials for solar dark photon absorption that builds on the existence of a theoretical upper limit on the rate at which dark photons can be absorbed in a terrestrial detector. The derivation of this limit relies on the so-called Kramers–Kronig relations -- consistency relations constraining any physical observable that can be expressed in terms of generalized susceptibilities~\cite{Catena:2024rym}, such as the solar dark photon absorption rate. As we will see later, these relations take different forms when applied to the longitudinal and transverse parts of the absorption rate. For applications of Kramers-Kronig relations in the context of DM direct detection, see \cite{Lasenby} and \cite{Catena:2025sxz}.

Depending on the dark photon mass, we find that an optimal detector material should have the following properties:
\begin{enumerate}
\item For the detection of dark photons with mass $m_V\ll\omega_p$, where $\omega_p$ is the detector's plasma frequency, the expected absorption rate of {\it longitudinal} dark photons in an optimal detector should approach the theoretical upper limit implied by the Kramers–Kronig relations;
\item For $m_V\approx\omega_p$, the absorption rate of {\it transverse} dark photons should exhibit a pronounced peak;
\item For $m_V\gg\omega_p$, {\it neither the longitudinal nor the transverse} rate should drop off too rapidly compared to the corresponding upper limits implied by the Kramers–Kronig relations.
\end{enumerate}
A non-trivial feature of our strategy is that it clarifies, for each mass regime, whether the expected longitudinal or transverse rate should guide the detector design, and it highlights the role of the detector's plasma frequency in this process.

The remainder of this work is organized as follows. In Sec.~\ref{sec:absorption}, we review the theory of dark photon production in the Sun and absorption in a terrestrial detector. In Sec.~\ref{sec:strategy}, we introduce our strategy to identify optimal detector materials for solar dark photon absorption, after reviewing the Kramers–Kronig relations and applying them separately to the longitudinal and transverse solar dark photon absorption rates. We conclude our analysis in Sec.~\ref{sec:conclusion} by applying this strategy to assess the performance of a selection of currently used detector materials, and with a brief discussion of metamaterials as detector materials for solar dark photon absorption.

\section{Production and absorption of dark photons}\label{sec:absorption}

In this section, we introduce the framework employed in our work to study the production and terrestrial detection of dark photons from the Sun.

\subsection{Dark photon model}\label{sec:model}
The Lagrangian for the specific dark photon model we consider in this work is given by~\cite{Pospelov1}
\begin{equation}
\mathcal{L} = -\frac{1}{4}F_{\mu\nu}F^{\mu\nu} - \frac{1}{4}V_{\mu\nu}V^{\mu\nu} - \frac{\kappa}{2}F_{\mu\nu}V^{\mu\nu} + \frac{m_V^2}{2}V_{\mu}V^{\mu} + e J_{\mathrm{em}}^{\mu}A_{\mu},
\end{equation}
where $F_{\mu\nu}$ and $V_{\mu\nu}$ are the field-strength tensors of the SM photon and the dark photon, respectively.
The kinetic-mixing term $-\frac{\kappa}{2} F_{\mu\nu}V^{\mu\nu}$ allows the dark photon to mix with the SM photon through the mixing parameter $\kappa$.
The mass term $\frac{m_V^2}{2}V_{\mu}V^{\mu}$ gives the dark photon a mass via the Stückelberg mechanism~\cite{Stueckelberg,Ruegg:2003ps} to which we restrict ourselves in the remainder of this work.
The last term in the Lagrangian corresponds to the familiar coupling of the SM photon to the QED current.

\subsection{Flux of dark photons from the Sun}\label{sec:flux}
\begin{figure}
    \centering
    \includegraphics[width=\linewidth]{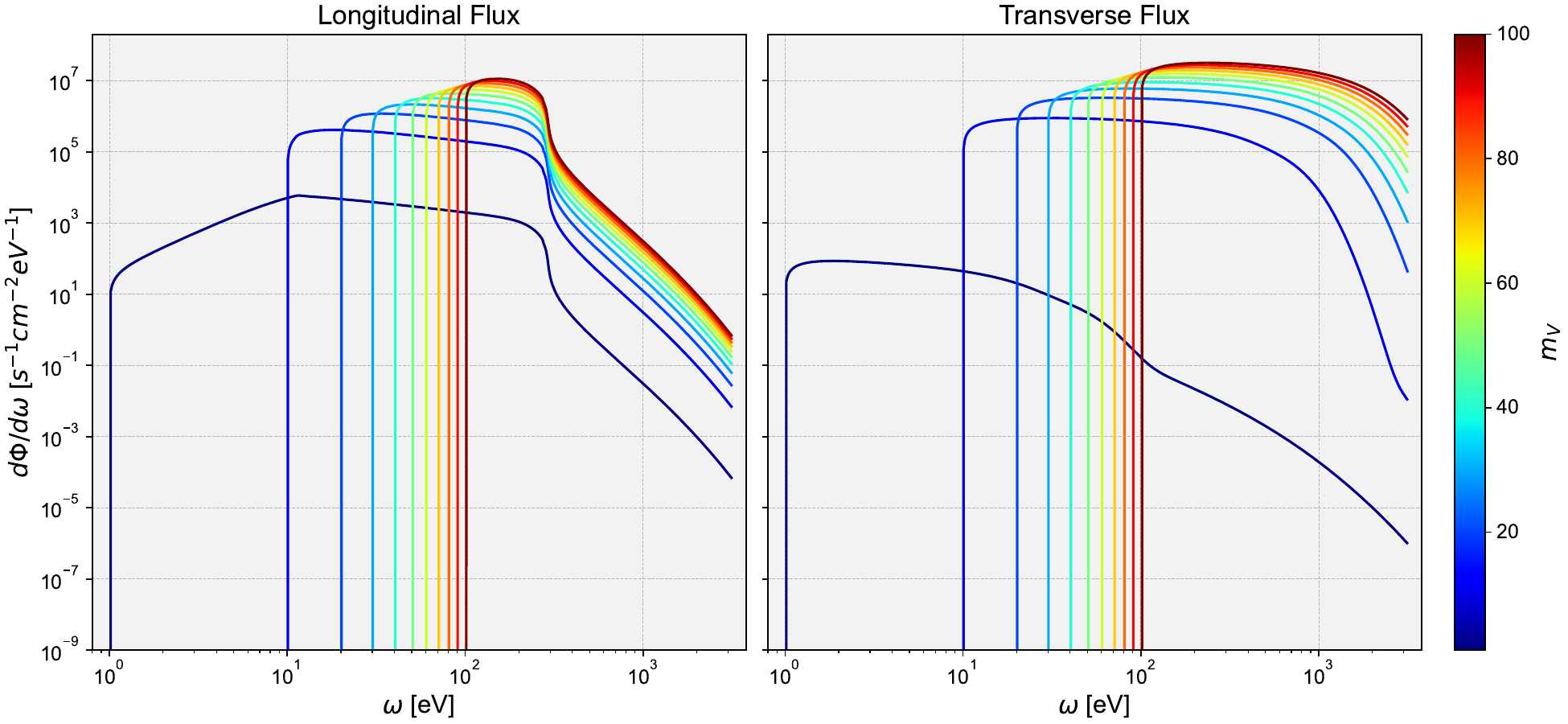}
    \caption{Flux of dark photons from the Sun, computed following Ref.~\cite{Pospelov2}, shown separately for the longitudinal and transverse components and plotted for dark photon masses between $1\text{ eV}$ and $100\text{ eV}$. The calculations are based on the solar model BS05(OP)~\cite{Bahcall_2005}, with $\kappa=10^{-16}$. For frequencies (dark photon masses) close to the plasma frequency in the solar medium, the production of longitudinal (transverse) dark photons is resonantly enhanced. However, for sufficiently large frequencies, the condition $\omega \sim \omega_p$ is never fulfilled, which explains the feature at $\omega\approx295\text{ eV}$ in the longitudinal flux \cite{Pospelov2}.
    }
    \label{fig:flux}
\end{figure}

Because of the kinetic mixing between the dark photon and the SM photon, astrophysical objects that emit large fluxes of SM photons -- such as the Sun -- are also expected to produce dark photons through this same mixing mechanism~\cite{Redondo_2008}. The production rate of dark photons in the solar interior has been computed in previous studies~\cite{Redondo_2008,Redondo_2013,Pospelov2}. Following these works, we evaluate the flux of solar dark photons reaching Earth under the assumption that, due to their extremely weak coupling, all produced dark photons freely escape the Sun without significant reabsorption or scattering. The resulting flux can be written in terms of the in-medium differential production rate of transverse and longitudinal dark photons, $d\Gamma^{Prod,V}_{T,L}/(d \omega dV)$, as

\begin{equation}
    \frac{d\Phi_{T,L}}{d\omega} = \frac{1}{(1 \text{AU})^2}\int_0^{R_\odot} r^2 dr \,  \frac{d\Gamma^{Prod,V}_{T,L}}{d\omega dV}\,,
    \label{eq:flux}
\end{equation}
where we assume that the solar properties entering the production rate depend only on the radial coordinate, and $R_\odot$ denotes the radius of the Sun. The rate $d\Gamma^{Prod,V}_{T,L}/(d \omega dV)$ can be factorized into a component, $d\Gamma^{Prod}_{T,L}/(d \omega dV)$, corresponding to the differential rate for the tree-level production of dark photons at $\kappa=1$, and an effective mixing parameter denoted by $\kappa_{T,L}$:
\begin{equation}
    \frac{d\Gamma^{Prod,V}_{T,L}}{d\omega dV} = \kappa_{T,L}^2\frac{d\Gamma^{Prod}_{T,L}}{d\omega dV}
    \label{eq:diff_prod}
\end{equation}
with
\begin{equation}
    \kappa_{T,L}^2=\frac{\kappa^2 m_V^4}{(m_V^2-\re \Pi_{T,L})^2+(\im \Pi_{T,L})^2}.
    \label{eq:eff_mixing}
\end{equation}
The effective mixing parameter encapsulates both the dark photon-SM photon vertex ($-i\kappa m_V^2$) and the in-medium propagator of the SM photon, given by \cite{Pospelov1,Pospelov2}
\begin{equation}
    \langle A_\mu,A_\nu\rangle = \frac{\eta_{\mu\nu}}{k^2-\Pi_{T,L}}=\frac{\eta_{\mu\nu}}{m_V^2-\Pi_{T,L}},
    \label{eq:photonProp}
\end{equation}
where the second equality comes from considering only on-shell dark photons.
The quantities $\Pi_{L}$ and $\Pi_{T}$ represent the longitudinal and transverse part of the in-medium SM photon polarization tensor $\Pi^{\mu\nu}$, namely
\begin{equation}
    \Pi^{\mu\nu} = \Pi_T \sum _{i=1,2} \epsilon_{i,T}^{\mu} \epsilon_{i,T}^{ \nu} + \Pi_L \epsilon_L^{\mu} \epsilon_L^{\nu}\,,
    \label{eq:poltensor_decom}
\end{equation}
where we assume an isotropic medium, while $\epsilon_{L}^{ \nu}$ and $\epsilon_{i,T}^{ \nu}$, $i=1,2$ are longitudinal and transverse polarization vectors. Interestingly, not only the effective mixing parameter $\kappa_{T,L}$, but the entire rate $d\Gamma^{Prod}_{T,L}/(d \omega dV)$ can be expressed in terms of $\Pi_{L}$ and $\Pi_{T}$. This connection arises from two key observations, which we briefly review below. First, at finite temperature $T\neq 0$, the imaginary part of $\Pi_{T,L}$ can be related to the total absorption rate $\Gamma_{T,L}^{abs, V}$ of in-medium transverse and longitudinal dark photons \cite{PhysRevD.28.2007},
\begin{equation}
    {\rm Im}\,\Pi_{T,L} = - \frac{\omega}{\kappa_{T,L}^2} \left(1- e^{-\omega/T}\right)\Gamma_{T,L}^{abs, V}\,.
    \label{eq:abs1}
\end{equation}
Second, by invoking detailed balance between dark photon absorption and production, and assuming that the dark photon distribution function is thermalized at temperature T, the absorption rate can be written as~\cite{Pospelov2},
\begin{align}
\Gamma_{T,L}^{abs,V} = \frac{2 \pi^2}{\omega \sqrt{\omega^2-m_V^2}} \frac{d\Gamma^{Prod, V}_{T,L}}{d\omega dV} \, e^{\omega/T} \,.
\label{eq:abs2}
\end{align}
Consequently, by combining Eqs.~\ref{eq:abs1} and \ref{eq:abs2}, one can rewrite the rate $d\Gamma^{Prod, V}_{T,L}/(d\omega dV)$ as follows
\begin{equation}
    \frac{d\Gamma^{Prod,V}_{T,L}}{d\omega dV} = \frac{\kappa_{T,L}^2}{e^{w/T}-1}\frac{\sqrt{\omega^2-m_V^2}}{2\pi^2}\left( - {\rm Im}\,\Pi_{T,L} \right)\,.
\end{equation}
The problem of computing the flux of dark photons from the Sun, $d\Phi_{T,L}/d\omega$, can thus be recast as the problem of determining the real and imaginary parts of the in-medium SM photon polarization tensor.
Following~\cite{Pospelov2}, and assuming that electrons in the Sun form a non-relativistic, non-degenerate gas, the real parts of $\Pi_{L}$ and $\Pi_{T}$ can be written as
\begin{subequations}\label{eq:Pi_omega}
\begin{align}
    \re\Pi_{L} &= \omega_p^2 \frac{m_V^2}{\omega^2},
    \label{eq:Pi_omega:L}\\
    \re\Pi_{T} &= \omega_p^2,
    \label{eq:Pi_omega:T}
\end{align}
\end{subequations}
where $\omega_p^2 = 4\pi \alpha n_e / m_e$, with $\alpha$ the fine-structure constant, $n_e$ the electron density in the Sun, and $m_e$ the electron mass. Note that $n_e$, and therefore $\omega_p$, are functions of the radial coordinate $r$. Below, we assume that bremsstrahlung is the dominant dark photon production channel and, following  \cite{Pospelov2} and \cite{Redondo_2008}, respectively,  for the imaginary parts of $\Pi_{L}$ and $\Pi_{T}$ we use,
\begin{align}
{\rm Im}\,\Pi_L = -\frac{32 \pi^2 \alpha^3 n_e }{3 m_e^2}  \frac{m_V^2}{\omega^4}
\sqrt{\frac{m_e}{2 \pi T}} 
 \left(e^{\omega/T}-1\right) e^{-\frac{\omega}{2 T}}
\sum_i Z_{i}^2n _{Z_i} K_0\left(\frac{\omega}{2 T}\right)
\end{align}
and 
\begin{align}
{\rm Im}\,\Pi_T = -\frac{32 \pi^2 \alpha^3 n_e }{3 m_e^2}  \frac{1}{\omega^2}
\sqrt{\frac{m_e}{2 \pi T}} 
 \left(e^{\omega/T}-1\right) e^{-\frac{\omega}{2 T}}
\sum_i Z_{i}^2n _{Z_i} K_0\left(\frac{\omega}{2 T}\right)
\end{align}
where $n_{Z_i}$ ($-i e Z_i$) is the density (charge) of the $i$-th ion in the Sun, while $K_0$ is a modified Bessel function of the second kind. In the numerical applications, we employ the solar model BS05(OP) \cite{Bahcall_2005} for the radial profiles of $\omega_p(r)$, $n_{Z_i}(r)$, $n_e(r)$, and the temperature $T(r)$. This choice allows for a direct comparison with the results reported in \cite{Pospelov2, Redondo_2008}. Fig.~\ref{fig:flux} shows the dark photon flux we find from Eq. \ref{eq:flux} for $\kappa=10^{-16}$ and a range of dark photon masses from 1 to 100 eV.

\subsection{Absorption rate in a detector}
After being emitted in the Sun, dark photons propagate freely through space and reach the Earth, where they can eventually be detected via absorption in underground detectors. Assuming that such detectors operate at temperatures much smaller than the dark photon mass, the rate of dark photon absorption in a terrestrial detector can be computed from Eq. \ref{eq:abs1} in the $T\rightarrow 0$ limit:
\begin{align}
\Gamma_{T,L}^{\text{abs},V} = -\kappa_{T,L}^2,\frac{{\rm Im} \,\Pi_{T,L}}{\omega}\,.
\label{eq:rate_earth}
\end{align}
Although Eq. \ref{eq:rate_earth} adopts the same notation as Eq. \ref{eq:abs1}, it is important to emphasise that in Eq. \ref{eq:rate_earth}, $\Gamma_{T,L}^{\text{abs},V}$ denotes the absorption rate of individual dark photons on Earth, not in the Sun. Consequently, in Eq. \ref{eq:rate_earth}, ${\rm Im}\,\Pi_{T,L}$ refers to the longitudinal and transverse components of the SM photon polarization tensor in the detector material, rather than in the solar medium.

Assuming an isotropic and non-magnetic detector material, and neglecting contributions from the diamagnetic current, the components of the polarization tensor can be written as follows \cite{Solyom, Trickle_2020}
\begin{subequations} \label{eq:dielectric_Pi}
\begin{align}
    \Pi_L&=q^2(1-\varepsilon_r),
    \label{eq:dielectric_Pi:L}\\
    \Pi_T &= \omega^2 (1-\varepsilon_r) \,.
    \label{eq:dielectric_Pi:T}
\end{align}
\end{subequations}
where $\varepsilon_r\equiv\varepsilon_r(\omega,\bm{q})$ is the dielectric function. Note that isotropy is a reasonable assumption in the case of dark photon absorption, where the momentum transfer involved in the process is negligible and the resulting electronic excitations propagate through the detector in the long-wavelength limit. In this regime, any medium effectively appears isotropic, since the wavelength associated with the momentum transfer in the absorption process is large compared to the characteristic scale of detector anisotropies. Combining Eqs. \ref{eq:dielectric_Pi} with Eq. \ref{eq:rate_earth}, one obtains,
\begin{subequations}\label{eq:absrate}
\begin{align}
    \Gamma^{abs,V}_{L} &=\frac{\kappa^2 m_V^2}{\omega}\im\left(\frac{-1}{\varepsilon_r}\right)\,,
    \label{eq:absrate:L}\\
    \Gamma^{abs,V}_{T} &=\frac{\kappa^2 m_V^4}{\omega}\im\left(\frac{-1}{(\varepsilon_r-1)\omega^2+m_V^2}\right)\,.
    \label{eq:absrate:T}
\end{align}
\end{subequations}
The quantity $\im \varepsilon_r^{-1}$ appearing in Eq.~\ref{eq:absrate:L} is known as the energy loss function (ELF), as it quantifies the energy lost by a particle traversing a material~\cite{Solyom}. Using the absorption rates in Eq.~\ref{eq:absrate} together with the dark photon fluxes in Eq.~\ref{eq:flux}, the rate of absorption events per unit time and detector volume can finally be expressed as~\cite{Pospelov1}
\begin{equation}
    n = \int_{\omega_{min}}^{\omega_{max}}\frac{\omega d\omega}{|\bm{q}|}\left( \frac{d\Phi_L}{d\omega}\Gamma^{abs,V}_L + \frac{d\Phi_T}{d\omega}\Gamma^{abs,V}_T\right) \,,
    \label{eq:eventrate}
\end{equation}
where $\omega_{min}=\max[\omega_{det},m_V]$, $\omega_{det}$ is the detector energy threshold, and $\omega_{max}$ is the maximum frequency that is experimentally observable.

\section{Strategy for optimal detector material identification}\label{sec:strategy}
Our strategy to identify optimal detector materials for solar dark photon absorption builds on the existence of a theoretical upper limit on the absorption rate. The derivation of this limit relies on consistency relations -- known as Kramers–Kronig relations -- which we briefly review below, before applying them separately to the longitudinal and transverse absorption rates.

\subsection{Kramers-Kronig relations}\label{sec:KramersKronig}
Let $\chi(\omega,\bm{q})$ be a function of $\omega$ that can be analytically continued into the upper half of the complex plane. If, in addition, $\chi(\omega,\bm{q}) \rightarrow 0$ for $|\omega|\rightarrow \infty$, then $\chi(\omega,\bm{q})$ satisfies the following relations
\begin{subequations}
\label{eq:KK}
\begin{align}
    \re\chi(\bm{q},\omega)&= \frac{1}{\pi}\mathcal{P}\int\limits_{-\infty}^\infty\frac{\im\chi(\bm{q},\omega')}{\omega'-\omega}d\omega', \\
    \im\chi(\bm{q},\omega)&= -\frac{1}{\pi}\mathcal{P}\int\limits_{-\infty}^\infty\frac{\re\chi(\bm{q},\omega')}{\omega'-\omega}d\omega',
\end{align}
\end{subequations}
known as Kramers–Kronig relations. Note that if $\omega$ ($\bm{q}$) is the frequency (momentum) conjugate to the time variable $t$ (the position vector ${\bm r}$), then the Fourier transform of $\chi(\bm{q},\omega)$, denoted by $\chi(\bm{r},t)$, is a causal function; that is, it is proportional to the step function $\theta(t)$. This condition is satisfied by any response function (or generalized susceptibility) that relates the strength of an external perturbation to the amplitude of the induced fluctuations of physical observables in an interacting electron system. Indeed, the response to a perturbation can manifest itself only after the perturbation has been applied, which ensures the causality of $\chi(\bm{r},t)$. When the perturbation is a real function, $\chi(\bm{r},t)$ is also real and its Fourier transform obeys
\begin{equation}
    \re \chi(\bm{q},\omega)=\re \chi(\bm{q},-\omega)\, , \quad \im \chi(\bm{q},\omega)= -\im \chi(\bm{q},-\omega).
\end{equation}
An important consequence of the Kramers–Kronig relations is that they imply the sum rule \cite{Solyom}
\begin{equation}
    \chi(\bm{q},0) = \frac{1}{\pi}\int_{-\infty}^\infty\frac{\im\chi(\bm{q},\omega)}{\omega}d\omega = \frac{2}{\pi}\int_{0}^\infty\frac{\im\chi(\bm{q},\omega)}{\omega}d\omega 
    \label{eq:sumrule}
\end{equation}
where the second equality follows from $\im \chi(\bm{q},\omega)$ being an odd function of $\omega$. Furthermore, when $\im\chi(\bm{q},\omega) \ge 0$ for $\omega$ in the range $[0, +\infty)$, Eq. \ref{eq:sumrule} leads to the inequality, 
\begin{equation}
     \int_{\omega_{min}}^{\omega_{max}}\frac{\im\chi(\bm{q},\omega)}{\omega}d\omega\leq\frac{\pi}{2}\chi(\bm{q},0)
\end{equation}
for finite, positive $\omega_{min}$ and $\omega_{max}$. One example of a response function is the dielectric function $\varepsilon_r(\bm{q},\omega)$, which relates the electric displacement field to the electric field strength. While $\varepsilon_r(\bm{q},\omega)$ is analytic in the upper half of the
complex plane (and its Fourier transform is therefore causal), it approaches unity in the limit $|\omega| \rightarrow \infty$. For this reason, the Kramers-Kronig relations apply not directly to $\varepsilon_r(\bm{q},\omega)$ but to $\varepsilon_r(\bm{q},\omega)-1$. Similarly, the Kramers-Kronig relations can be applied to the imaginary part of the inverse dielectric function, the ELF $\im (-\varepsilon_r^{-1}(\bm{q},\omega))$ \cite{Solyom}. Since the ELF is always larger than or equal to zero for a physical system in its ground state, in addition to the Kramers-Kronig relations it also satisfies
\begin{equation}
    \int_{\omega_{min}}^{\omega_{max}}\frac{d\omega}{\omega}\im\left(\frac{-1}{\varepsilon_r(\bm{q},\omega)}\right)\leq \frac{\pi}{2}\left( 1-\frac{1}{\varepsilon_r(\bm{q},0)}\right).
    \label{eq:sumrule_elf}
\end{equation}
Interestingly, the Kramers-Kronig relations can also be formulated for functions that are analytic in the lower half of the complex plane. In this case, the right-hand side in the Kramers-Kronig relations, Eq.~\ref{eq:KK},  and in the related sum rule, Eq.~\ref{eq:sumrule}, have to be multiplied by (-1). Explicitly, 
\begin{equation}
    \int_0^\infty \frac{d\omega}{\omega} \im\chi(\bm{q},\omega) = \pm\frac{\pi}{2}\chi(\bm{q},0)
    \label{eq:sumrule_general}
\end{equation} for $\chi(\omega, \bm{q})$ fulfilling the following requirements
\begin{enumerate}
    \item $\lim\limits_{|\omega|\rightarrow\infty}\chi(\bm{q},\omega)=0$,
    \item $\chi(\bm{q},\omega)$ can be analytically continued into the upper half plane ($+$) or into the lower half plane ($-$),
    \item $\im \chi(\bm{q},\omega)$ is an odd function of $\omega$.
\end{enumerate}

\subsection{Derivation of an upper limit on the absorption rate}\label{sec:upperlim}
In this subsection, we derive a theoretical upper limit on the absorption rate of solar dark photons by applying the Kramers-Kronig relations separately to the longitudinal and transverse rates.
\subsubsection{Longitudinal rate}\label{sec:long}
Retaining only the longitudinal contribution to the event rate in Eq.~\ref{eq:eventrate} yields
\begin{equation}
    n_L = \int_{\omega_{min}}^{\omega_{max}} \frac{d\omega}{\omega} \im\left(\frac{1}{\varepsilon_r(\bm{q},\omega)}\right) \kappa^2 m_V^2 \frac{\omega}{\sqrt{\omega^2-m_V^2}} \frac{d\Phi_L}{d\omega}\,.
    \label{eq:eventrateL}
\end{equation}
In this expression, the integral appearing in Eq.~\ref{eq:sumrule_elf} can be directly identified. This immediately leads to an upper limit on the longitudinal rate:
\begin{align}
    n_L &\leq \kappa^2 m_V^2\max\limits_{\omega\in[\omega_{min},\omega_{max}]} \left( \frac{\omega}{\sqrt{\omega^2-m_V^2}} \frac{d\Phi_L}{d\omega} \right) \int_{\omega_{min}}^{\omega_{max}} \frac{d\omega}{\omega} \im\left(\frac{1}{\varepsilon_r(\bm{q},\omega)}\right) \nonumber\\
    &\leq \kappa^2 m_V^2\max\limits_{\omega\in[\omega_{min},\omega_{max}]} \left( \frac{\omega}{\sqrt{\omega^2-m_V^2}} \frac{d\Phi_L}{d\omega} \right)  \frac{\pi}{2}\left( 1-\frac{1}{\varepsilon_r(\bm{q},0)}\right).
    \label{eq:limit_L}
\end{align}
This result is analogous to the bound derived in Ref.~\cite{Lasenby}, where Eq.~\ref{eq:sumrule_elf} is applied in the context of DM scattering with electrons. 

Eq.~\ref{eq:limit_L} still retains an explicit material dependence through $\varepsilon_r^{-1}$. A fully material-independent bound follows from the fact that dark photon absorption occurs in the limit $|\bm{q}|\approx 0$, which is the relevant case here because the momentum transfer in the absorption process is negligible compared to the electron mass. In this regime, $\varepsilon_r(\bm{q},0) \rightarrow +\infty$ for metals, while for semiconductors it remains finite but greater than unity \cite{Solyom}. Consequently, a universal upper limit is
\begin{equation}
    n_{L,opt}= \kappa^2 m_V^2 \frac{\pi}{2}\max\limits_{\omega\in[\omega_{min},\omega_{max}]} \left( \frac{\omega}{\sqrt{\omega^2-m_V^2}} \frac{d\Phi_L}{d\omega} \right).
    \label{eq:nopt_L}
\end{equation}

\subsubsection{Transverse rate}\label{sec:trans}
Setting $|{\bm q}|$ to zero and introducing the notation $\varepsilon_r(\omega) \equiv \varepsilon({\bm q}, \omega)$, the transverse contribution to the event rate in Eq.~\ref{eq:eventrate} is
\begin{equation}
    n_T = \int_{\omega_{min}}^{\omega_{max}}\frac{ d\omega}{\omega}\im\left(\frac{-1}{(\varepsilon_r(\omega)-1)\omega^2+m_V^2}\right)\kappa^2 m_V^4\frac{\omega}{\sqrt{\omega^2-m_V^2}} \frac{d\Phi_T}{d\omega}.
    \label{eq:eventrateT}
\end{equation}
This expression also contains an integral over the imaginary part of a function. Unlike in the longitudinal case, however, it is not immediately clear whether the Kramers-Kronig relations -- and the resulting sum rule in Eq.~\ref{eq:sumrule_general} -- can be directly applied. We therefore need to verify that the function
\begin{equation}
    \chi(\omega) \equiv \frac{-1}{(\varepsilon_r(\omega)-1)\omega^2+m_V^2}
\end{equation}
satisfies the conditions required for Eq.~\ref{eq:sumrule_general}. To check the first condition, we evaluate the $|\omega| \rightarrow \infty$ limit:
\begin{equation}
    \lim\limits_{|\omega|\rightarrow\infty} \frac{-1}{(\varepsilon_r(\omega)-1)\omega^2+m_V^2} = \lim\limits_{|\omega|\rightarrow\infty} \frac{-1}{-\frac{\omega_p^2}{\omega^2}\omega^2+m_V^2} = \frac{1}{\omega_p^2-m_V^2},
\end{equation}
where we used the large $\omega$-behaviour $\epsilon_r(\omega)= 1 - \omega_p^2/\omega^2$ \cite{Solyom}. Thus, to enforce the required falloff we introduce the shifted function
\begin{equation}
    \chi_{\text{new}}(\omega) \equiv\frac{-1}{(\varepsilon_r(\omega)-1)\omega^2+m_V^2} - \frac{1}{\omega_p^2-m_V^2},
\end{equation}
which vanishes as $|\omega| \rightarrow \infty$ and has the same imaginary part as $\chi(\omega)$. Next, we check the analyticity condition by finding the poles of $\chi_{\text{new}}(\omega)$.
They correspond to the solutions of
\begin{equation}
    (\varepsilon_r(\omega)-1)\omega^2+m_V^2 = 0.
    \label{eq:condition2}
\end{equation}
An explicit analytic form for $\varepsilon_r(\omega)$ is therefore required. A convenient choice is the semi-classical Drude–Lorentz oscillator model \cite{Echenique}:
\begin{align}
    \varepsilon_r(\bm{q},\omega)=1+\frac{\omega_p^2}{E_g^2+\frac{3}{5}v_F^2|\bm{q}|^2+\frac{|\bm{q}|^4}{4m_e^2}-\omega(\omega+i\gamma)},
\end{align}
where $\omega_p$ is the plasma frequency of the material, $E_g$ the band gap, $v_F$ is the Fermi velocity given by $v_F=(3\pi^2 n_e)^{1/3}/m_e=(3\pi\omega_p^2/4\alpha m_e^2)^{1/3}$ and $\gamma$ is a damping rate. The term proportional to $v_F^2$ describes plasmon dispersion \cite{Raether}, whereas the term depending on $|{\bm q}|^4$ accounts for single-particle excitations \cite{Echenique}. In the regime $|\bm{q}|^2\ll\omega m_e$, relevant for absorption of light dark photons, the ${\bm q}$-dependence becomes negligible, and the dielectric function simplifies to
\begin{equation}
    \varepsilon_r(\omega)=1+\frac{\omega_p^2}{E_g^2-\omega(\omega+i\gamma)}.
    \label{eq:epsilonDrude}
\end{equation}
The well-known Mermin model yields the same expression for $|{\bm q}|=0$ \cite{Vos_Grande}. Substituting Eq.~\ref{eq:epsilonDrude} into Eq.~\ref{eq:condition2} gives
\begin{align}
    \omega_\pm &= \frac{-i\gamma m_V^2}{2(m_V^2-\omega_p^2)}\pm\sqrt{\frac{-\gamma^2 m_V^4}{4(m_V^2-\omega_p^2)^2}+\frac{m_V^2E_g^2}{m_V^2-\omega_p^2}}.
\end{align}
For $E_g=0$, which is the case for metals, the solutions reduce to $\omega_+=0$ and $\omega_-=-i\gamma m_V^2/(m_V^2-\omega_p^2)$, which lies in the upper half-plane for $m_V^2<\omega_p^2$ and in the lower half-plane for $m_V^2>\omega_p^2$. For $E_g\neq0$ and $m_V^2>\omega_p^2$, the solution $\omega_-$ always lies in the lower half plane. A pole at $\omega_+$ in the upper half plane would require
\begin{align}
    \frac{\gamma m_V^2}{2(m_V^2-\omega_p^2)}&<\sqrt{\frac{\gamma^2 m_V^4}{4(m_V^2-\omega_p^2)^2}-\frac{m_V^2E_g^2}{m_V^2-\omega_p^2}}  \quad\text{and}\quad \frac{\gamma^2 m_V^4}{4(m_V^2-\omega_p^2)^2}>\frac{m_V^2E_g^2}{m_V^2-\omega_p^2}%\nonumber\\\nonumber\\
    %\Rightarrow E_g^2&<0
\end{align}
and therefore $E_g^2<0$ which is unphysical. Consequently, $\omega_+$ must also lie in the lower half-plane. The case $E_g\neq0$ and $m_V^2<\omega_p^2$ proceeds analogously, with $\omega_+$ and $\omega_-$ always lying in the upper half plane for $E_g^2>0$. Finally, we verify that 
$\im \chi_{\text{new}}(\omega)$ is odd in $\omega$. Using Eq.~\ref{eq:epsilonDrude},
\begin{align}
    \im \chi_{\text{new}}(\omega) &= \im\left( \frac{-1}{\left(\frac{\omega_p^2}{E_g^2-\omega(\omega+i\gamma)}\right)\omega^2+m_V^2} \right)\nonumber\\
    &= \frac{\omega_p^2\omega^3\gamma}{\left( \omega_p^2\omega^2+m_V^2(E_g^2-\omega^2) \right)^2+(m_V^2\gamma\omega)^2} = -\im\chi_{\text{new}}(-\omega).
\end{align}
Thus, all conditions for applying the sum rule Eq.~\ref{eq:sumrule_general} are satisfied: the ``$+$'' version applies for $m_V^2>\omega_p^2$, and the version with an additional ``$-$'' for $m_V^2<\omega_p^2$. To use the sum rule to bound the integral in Eq.~\ref{eq:eventrateT}, we must also ensure that $\im \chi_{\textrm new}>0$ for $\omega>0$. Rewriting,
\begin{equation}
\im\left(\frac{-1}{\Delta\varepsilon_r\omega^2+m_V^2}\right) = \frac{\im \left((\varepsilon_r(\omega)-1)\omega^2 + m_V^2 \right)}{\left|(\varepsilon_r(\omega)-1)\omega^2 + m_V^2\right|^2} = \frac{\im \varepsilon_r}{\left|(\varepsilon_r(\omega)-1)\omega^2 + m_V^2\right|^2}.
\end{equation}
The denominator is manifestly positive, and the numerator is positive because the ELF, $\im\left(-\varepsilon_r^{-1}\right)=\im\varepsilon_r/|\varepsilon_r|^2$, is positive. We can therefore bound Eq.~\ref{eq:eventrateT} as
\begin{align}
n_T \leq \kappa^2 m_V^2 \max\limits_{\omega\in[\omega_{min},\omega_{max}]} \left( \frac{\omega}{\sqrt{\omega^2-m_V^2}} \frac{d\Phi_T}{d\omega} \right) \frac{\pi}{2} \frac{1}{\left|\frac{m_V^2}{\omega_p^2}-1\right|}.
\label{eq:nopt_T_w}
\end{align}
Because the result depends explicitly on $\omega_p$, this is not a universal material-independent bound. In contrast to the longitudinal case, a general universal limit cannot be obtained. However, two regimes do allow a material-independent expression, leading to
\begin{equation}
n_{T,opt}= \kappa^2 m_V^2 \frac{\pi}{2} \max\limits_{\omega\in[\omega_{min},\omega_{max}]} \left( \frac{\omega}{\sqrt{\omega^2-m_V^2}} \frac{d\Phi_T}{d\omega} \right) \quad\text{for $m_V^2\ll\omega_p^2$ or $m_V^2\geq2\omega_p^2$}.
\label{eq:nopt_T}
\end{equation}
We remark that the analytic dielectric function in Eq.~\ref{eq:epsilonDrude} is an idealized model. It assumes a single plasmon pole at $\omega_p$, whereas real materials exhibit a distribution of plasma frequencies due to spatial variations in electron density. A more accurate description involves a weighted sum over oscillators with different $\omega_{p,i}$ and $\gamma_i$ \cite{Vos_Grande}:
\begin{equation}
\varepsilon_r(\omega)=1+\sum_i \frac{A_i \omega_{p,i}^2}{E_g^2-\omega(\omega+i\gamma_i)}.
\end{equation}
Using this more realistic form would prevent a closed-form analytical derivation of the bound, but the simplified model still captures the key material properties that control the absorption process.

\subsubsection{Comparison with material-specific rates}\label{sec:materials}

A key preparatory step in formulating our strategy for optimal material identification in solar dark photon absorption is to compare the event rate for a selection of real detector materials with the upper limits derived above. To perform this comparison, the integrands in Eqs.~\ref{eq:eventrateL} and \ref{eq:eventrateT} need to be evaluated for explicit choices of $\varepsilon_r$. In this work, we adopt parametrizations of experimental data based on a Drude–Lorentz oscillator model~\cite{Sun, Novak} to compute the ELF. We then use the Kramers–Kronig relations to obtain the real part of $\varepsilon_r^{-1}$ from the ELF, and thus reconstruct $\varepsilon_r$. For the integration boundaries in Eqs.~\ref{eq:eventrateL} and \ref{eq:eventrateT}, we take $\omega_{min}=m_V$, since $m_V$ is the smallest energy an on-shell dark photon can carry, and we choose $\omega_{max}$ to be the largest energy available in the parametrizations of \cite{Sun, Novak}. Because the solar dark photon flux decreases rapidly at high energies, our results are not very sensitive to the precise value of $\omega_{max}$. At the same time, the condition $\omega_{min}=m_V$ introduces an additional implicit mass dependence in the observable absorption rates. Finally, we note that the material-specific rates in Eqs.~\ref{eq:eventrateL} and \ref{eq:eventrateT} and the corresponding upper limits in eq.~\ref{eq:nopt_L} and \ref{eq:nopt_T} share the same dependence on $\kappa$. Their comparison is therefore insensitive to the specific choice of $\kappa$.

\begin{figure}
    \centering
    \includegraphics[width=\linewidth]{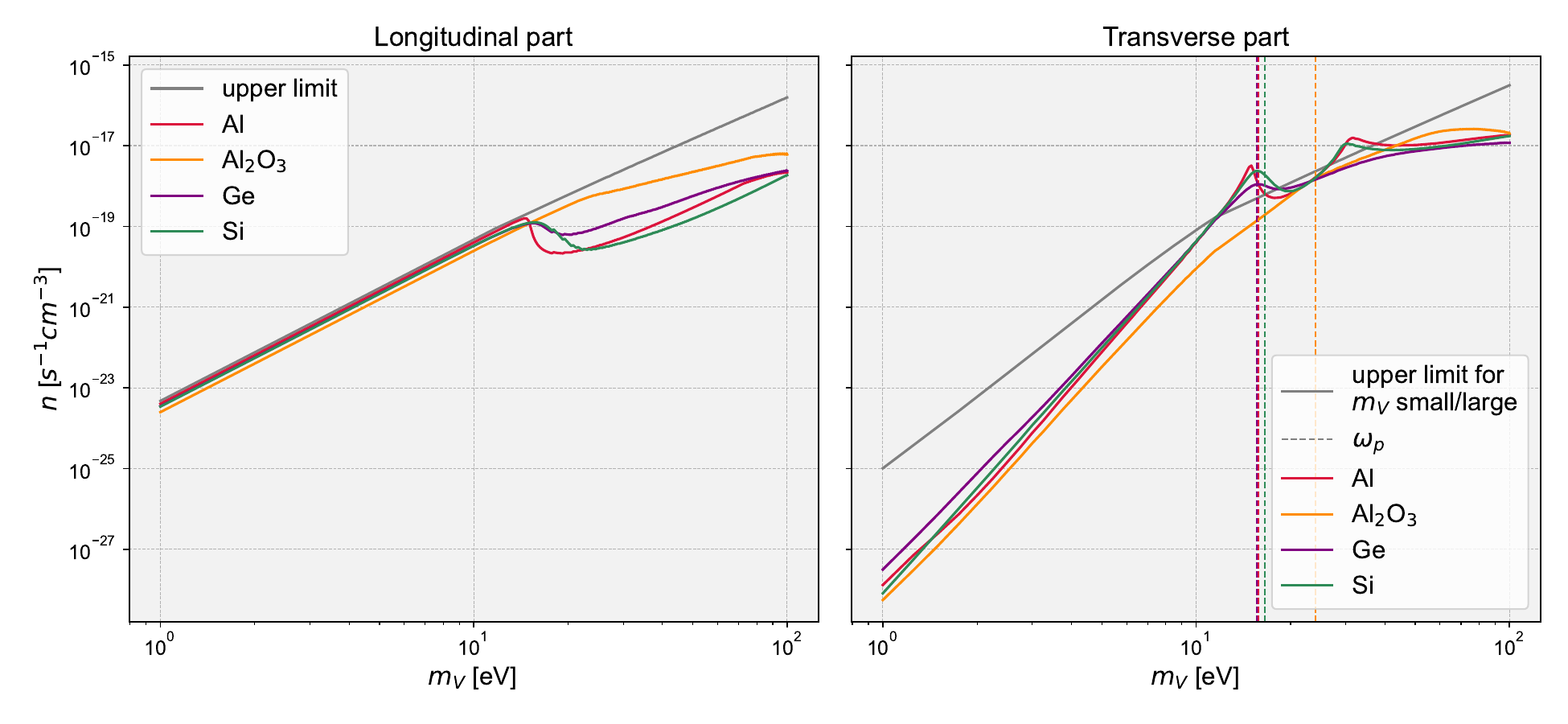}
    \caption{Comparison of material-independent upper limits with material-specific rates for four selected materials. $\kappa$ was set to $10^{-16}$ based on \cite{XENON_Collab_2025} but only affects absolute numbers not the relative comparison between rates and upper limits. For the transverse part also the plasma frequency of the material is indicated, where the material-independent upper limit is not valid.}
    \label{fig:materials}
\end{figure}
Fig.~\ref{fig:materials} shows the material-independent upper limits together with the material-specific rates for four representative materials. We include germanium (Ge), silicon (Si), and sapphire (Al$_2$O$_3$), as these are commonly used in direct detection experiments \cite{SuperCDMS_2018,SuperCDMS_2024,CRESST}. Since these are semiconductors or insulators, we also consider aluminium (Al) as an example of a metal. As anticipated from the results of the previous sections, the material-independent upper limit is generally respected for the longitudinal component but not for the transverse one. Near the plasma frequency, the transverse material-specific rates exhibit a peak that exceeds the material-independent limit. This behaviour is expected from the material-dependent bound in Eq.~\ref{eq:nopt_T_w}, which contains a pole at the plasma frequency. Both Si and Al display a secondary peak, while Al$_2$O$_3$ shows no clear peak. This indicates that the dielectric function model adopted for the calculation of the upper limit is not sufficiently accurate to reproduce all features of real materials. For this reason, the conditions under which the transverse material-independent limit applies must be relaxed to the regimes $m_V^2\ll\omega_p^2$ or $m_V^2\gg\omega_p^2$.

As one can see from Fig.~\ref{fig:materials}, the longitudinal material-specific rates approach their corresponding upper limit at low masses, up to approximately the plasma frequency. The mass dependence of the longitudinal upper limit follows a power law $m_V^4$, as expected: the absorption rate scales as $m_V^2$ (see Eq.~\ref{eq:absrate:L}), while the flux scales as $m_V^2 \sqrt{\omega^2 - m_V^2}$ (see Eqs.~\ref{eq:flux} and \ref{eq:Pi_omega:L}), where the square root cancels the factor $|\bm{q}|^{-1} = (\omega^2 - m_V^2)^{-1/2}$ appearing in the event rate integral Eq.~\ref{eq:eventrate}. The integrated material-specific rates acquire an additional mass dependence from the lower integration boundary. Owing to the shape of the ELF -- with a peak near the plasma frequency and very small values for energies below it \cite{Sun} -- this additional $m_V$ dependence becomes relevant only around the plasma frequency. In this region, the material-specific rates begin to deviate from the common $m_V^4$ scaling they share with the upper limit at lower masses. Materials with a very narrow ELF peak, such as Al, exhibit a sharp drop at the plasma frequency, whereas materials with broader peaks, such as Al$_2$O$_3$, deviate from the upper limit more gradually.

The mass dependence of the transverse upper limit and of the corresponding material-specific rates is considerably more intricate, because it does not factor out of the flux integral in Eq.~\ref{eq:flux}. As a consequence, the transverse upper limit and the material-specific transverse rates exhibit different mass scalings, and the latter do not approach the optimal limit as closely as in the longitudinal case. Nevertheless, the largest overall event rates are attained in the transverse channel near the plasma frequency, precisely where the material-independent upper limit ceases to apply.

\subsection{Strategy formulation}
\begin{figure}
    \centering
    \includegraphics[width=0.8\linewidth]{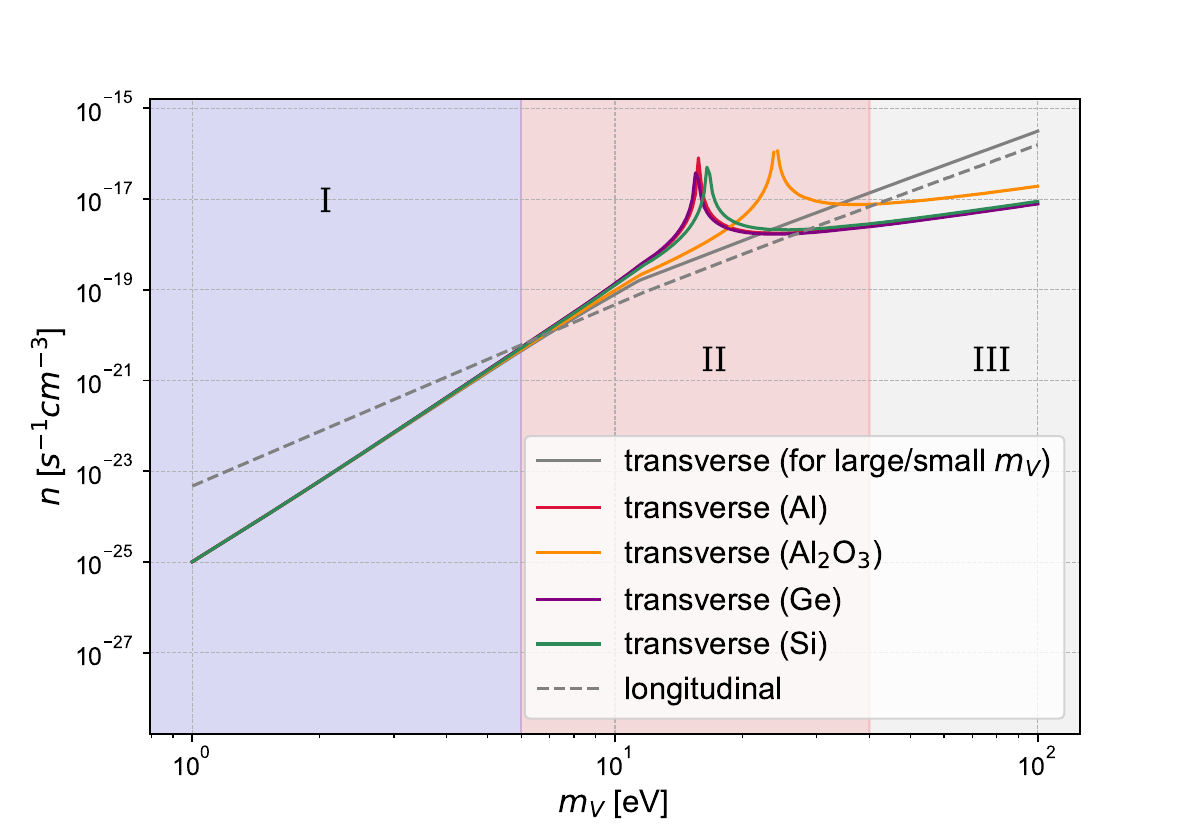}
    \caption{Comparison of material-independent upper limits for longitudinal and transverse part and the material-dependent upper limits for the transverse part. One can see three regions: I the longitudinal upper limit is higher, II the peaks at the plasma frequency in the material-dependent transverse upper limits dominate, III it is not obvious which contribution dominates.}
    \label{fig:limits}
\end{figure}
To formulate a strategy for assessing how suitable a detector material is for the absorption of solar dark photons, we summarize the results of the previous subsections in a single figure,  Fig.~\ref{fig:limits}. This figure shows the longitudinal and transverse material-independent upper limits, as well as the transverse material-specific absorption rates for the example materials introduced in the previous subsection. We do not include in the figure the longitudinal material-specific absorption rates because the longitudinal material-specific rates and the material-independent longitudinal upper limit almost coincide for $m_V\ll\omega_p$, as the analysis in Sec.~\ref{sec:trans} shows. Fig.~\ref{fig:limits} allows us to identify three distinct mass regimes: 1) For $m_V\ll\omega_p$, the longitudinal upper limit lies above the transverse one. 2) Near the plasma frequency, pronounced peaks in the material-specific transverse absorption rate are clearly visible; these exceed the corresponding material-independent upper limit, which is not valid in this mass region. 3) Finally, for $m_V\gg\omega_p$, the transverse material-independent upper limit becomes the dominant one, {\it i.e.} it is more stringent than the longitudinal material-independent upper limit. In this latter mass regime, the situation becomes more material-dependent: neither the longitudinal nor the transverse component clearly dominates in general, because in this regime the material-specific rates deviate more significantly from their respective upper limits (see Fig.~\ref{fig:materials}).

Based on these three regimes, a general strategy for evaluating the suitability of a material for dark photon absorption can be formulated:
\begin{enumerate}
\item For $m_V\ll\omega_p$, the {\it longitudinal} rate expected in an optimal detector material should approach the corresponding theoretical upper limit. 
\item For $m_V\approx\omega_p$, the {\it transverse} rate should exhibit a clear peak.
\item For $m_V\gg\omega_p$, {\it neither rate} should decrease too rapidly compared to the corresponding theoretical upper limits.
\end{enumerate}
This strategy can be directly applied to the example materials shown in Fig.~\ref{fig:materials}. Ge and Si perform similarly well, exhibiting high longitudinal rates at low masses and a clear transverse peak near the plasma frequency. Al$_2$O$_3$ displays the lowest longitudinal rate at low masses among the four materials and also lacks a pronounced transverse peak, making it less suitable than Ge and Si. Al, by contrast, has the highest longitudinal rate at low masses and a strong transverse peak. Although its longitudinal rate drops more sharply at the plasma frequency than in the other materials, this has little impact on the overall absorption rate because the transverse contribution dominates in this region. At higher masses, Al performs comparably to Si and Ge. We have performed analogous comparisons for other detector materials, including GaAs, ZnO, ZnS, Diamond, and SiO$_2$, confirming the validity of the general strategy outlined above.

\section{Summary and outlook}\label{sec:conclusion}
We have formulated a strategy to identify optimal detector materials for solar dark photon absorption. Our strategy builds
on the existence of a theoretical upper limit on the rate of solar dark photon absorption in terrestrial detectors. We have derived this upper limit
by applying the Kramers–Kronig relations to the longitudinal and transverse absorption rates separately.

Our strategy consists in optimizing the choice and design of a detector  material for solar dark photon absorption based on three criteria: For $m_V\ll\omega_p$, the expected longitudinal rate should approach the corresponding theoretical upper limit. For $m_V\approx\omega_p$, the transverse rate should exhibit a pronounced peak. Finally, for $m_V\gg\omega_p$, both the longitudinal and transverse rates should not decrease too rapidly compared to the corresponding theoretical upper limits. 

We have applied this strategy to assess the performance of a selection of potential detector materials, including Ge, Si, Al$_2$O$_3$ and Al. For example, we have found that, among these four materials, Al$_2$O$_3$ has the lowest longitudinal rate at low masses and does not have a clear peak in the transverse rate either. Consequently, it is less suitable for dark photon absorption than Ge or Si.

Finally, our strategy indicates that detection is most efficient when the material's plasma frequency matches the dark photon mass. If the plasma frequency could be tuned, one could scan different masses with maximal sensitivity. Metamaterials -- such as wire arrays, Josephson-junction arrays, or metal-dielectric heterostructures -- offer this tunability, though current implementations reach plasma frequencies only up to 100~$\mu$eV or so, which is too low for the mass range relevant here \cite{PhysRevLett.123.141802,PhysRevB.83.014511}. Moreover, it remains uncertain whether the assumptions about the dielectric function used to derive the upper limit apply to metamaterials. A full characterization of their dielectric properties is therefore required, and further research is needed to assess their viability for dark photon absorption experiments.

\acknowledgments
We would like to thank Jan Heisig for insightful discussions that helped improving the quality of the research presented here, and Julia Wiktor for providing us with useful references and clarifications concerning the modelling of the dielectric function. R.~C. acknowledges support from an individual research
grant from the Swedish Research Council (Dnr. 2022-04299) and from the Knut and Alice Wallenberg Foundation via the ``Light Dark Matter'' project (Dnr. KAW 2019.0080). R.~C. and T.~M.~B. also acknowledge support from the Swiss National Science Foundation (SNF), Beitrag Nr. 10.002.603. The research of MK is supported by the German Research Foundation DFG under grant 396021762 -- TRR 257: Particle Physics Phenomenology after the Higgs
Discovery.


% \begin{thebibliography}{99}
\bibliographystyle{unsrt}
\bibliography{references}

% \bibitem{Pospelov1}
% H. An, M. Pospelov, and J.Pradler, \emph{Dark Matter Detectors as Dark Photon He-
% lioscopes}, \emph{Phys. Rev. Lett.} {\bf 111} (2013)

% \bibitem{Pospelov2}
% H. An, M. Pospelov, and J.Pradler, \emph{New stellar constraints on dark photons}, \emph{Phys. Lett. B} {\bf 725} (2013)

% \bibitem{a}
% Author, \emph{Title}, \emph{J. Abbrev.} {\bf vol} (year) pg.

% \bibitem{b}
% Author, \emph{Title},
% arxiv:1234.5678.

% \bibitem{c}
% Author, \emph{Title},
% Publisher (year).

% Also, please have only one work for each \bibitem.

% \end{thebibliography}
\end{document}